\documentclass[aps,prd, twocolumn,groupedaddress,amsmath,amssymb,preprintnumbers,nofootinbib]{revtex4-2}

\usepackage[dvipdfmx]{graphicx}
\usepackage{dcolumn}
\usepackage{bm}
\usepackage{braket}
\usepackage{xcolor}
\usepackage{hyperref}

\begin{document}

\title{Reflection polarization of close binaries as a probe of axion dark matter birefringence}
\author{Tomoki Matsuoka}
\email{tmatsuoka0124@g.ecc.u-tokyo.ac.jp}
\affiliation{Department of Earth Science and Astronomy, Graduate School of Arts and Sciences, The University of Tokyo, Tokyo 153-8902, Japan}
\affiliation{Department of Physics, National Chung Hsing University, No. 145, Xingda Rd., South Dist., Taichung, 40227, Taiwan}
\author{Kimihiro Nomura}
\email{k.nomura@tap.scphys.kyoto-u.ac.jp}
\affiliation{Department of Physics, Kyoto University, Kyoto 606-8502, Japan}
\author{Hidetoshi Omiya}
\email{omiya@tap.scphys.kyoto-u.ac.jp}
\affiliation{Department of Physics, Kyoto University, Kyoto 606-8502, Japan}
\begin{abstract}
We propose close binary polarimetry as a probe of birefringence induced by ultralight axion dark matter.
In a close binary, reflection or scattering can generate a small linear polarization whose time dependence is locked to the orbital phase. This phase-locked polarization provides a template against which an oscillatory rotation of the polarization angle induced by the axion can be searched for.
We show that axion birefringence appears as sidebands around the orbital
harmonics. For a single bright binary, with parameters motivated by observed systems and current high-precision optical polarimetry, we estimate the sensitivity to the axion-photon coupling under white noise assumption to be the level of $10^{-12}$ GeV$^{-1}$ at an axion mass of $10^{-20}$ eV.
A future array of suitable binaries could further improve the sensitivity to $10^{-13}$ GeV$^{-1}$ in an optimistic scenario. This method could provide a complementary high-cadence optical probe of axion birefringence, compared to existing astrophysical searches.
\end{abstract}
\maketitle

\section{Introduction}

Light pseudoscalar fields, such as axions, are well motivated in physics beyond
the Standard Model. The QCD axion was originally introduced as a dynamical
solution to the strong CP problem, while more general axion-like particles arise
in many ultraviolet extensions of the Standard Model~\cite{
Peccei:1977hh,Weinberg:1977ma,Wilczek:1977pj,Preskill:1982cy,
Abbott:1982af,Dine:1982ah,Arvanitaki:2009fg,Svrcek:2006yi}.
In particular, ultralight axions are promising dark matter (DM) candidates
\cite{Hui:2016ltb}. Such DM is well described as a
classical oscillating field with an oscillation frequency set by its mass
$\mu$. Searches for ultralight axion dark matter can therefore be formulated as searches for weak, coherent, time-dependent signals.

A particularly direct signature is the birefringence of light induced by the
axion-photon coupling~\cite{Harari:1992ea, Carroll:1989vb}. In a time-dependent axion background, the left- and right-circular
polarization modes of the photon acquire different phases, resulting in a
rotation of the linear polarization angle. In the geometric optics limit, this
rotation is determined by the difference between the axion field values at the observation and emission events. For a coherent ultralight
axion field, the signal is therefore an achromatic, oscillatory polarization
rotation with a frequency determined by the axion mass $\mu$. The observational requirement is a linearly polarized source whose intrinsic polarization angle
can be predicted, modeled, or reconstructed from the data.

Several classes of polarized astrophysical and cosmological sources have been used in searches for
axion-induced birefringence, including the cosmic microwave background
(CMB)~\cite{Fedderke:2019,Ferguson:2022,BICEP:2021}, protoplanetary disks
\cite{Fujita:2018zaj,Narita:2026dvw}, and pulsars
\cite{Liu:2021zlt,Castillo:2022,EPTA:2024gxu,Yuwen:2026zjk} (see~\cite{Chen:2019fsq, Wang:2024sdz, Huang:2025rrb, POLARBEAR:2025djl} for other classes of astrophysical objects). For the CMB, one
searches for an oscillatory rotation of the CMB polarization that is coherent
across the observed sky. For protoplanetary disks, light scattered by the disk is expected to have a characteristic spatial pattern of linear polarization, which can be used as a template. Pulsars emit linearly polarized radio pulses, and repeated observations can reveal an oscillatory polarization rotation induced by axion DM. Moreover, combining observations of many pulsars can help separate the common axion-induced contribution from pulsar-dependent terms.

In this work, we propose close binary stars as a new class of polarized sources for probing axion DM through birefringence.
In a close binary, radiation from one component can be scattered in the
atmosphere of its companion, producing a small but measurable linear
polarization. For hot stars, the dominant mechanism is Thomson scattering, and
the resulting polarization is controlled by the orbital geometry. The time
dependence of the polarization is therefore locked to the binary phase and can
be expanded in harmonics of the orbital frequency $\Omega_{\rm orb}$
\cite{1978A&A....68..415B,1978ApJ...221..200R}. The reflection induced polarization
has been observed in systems such as Spica and $\mu^1$ Sco, with polarization
fractions of order $10^{-4}$--$10^{-3}$. Moreover, polarized radiative-transfer
models reproduce the observed phase dependence
\cite{2019NatAs...3..636B,cotton2020phase}. Close binaries can therefore
provide deterministic time-domain polarization templates.

We show that axion-induced birefringence produces a characteristic modulation
of the phase-locked polarization from a close binary. If the intrinsic Stokes
parameters contain harmonics at $n\Omega_{\rm orb}$, with $n=1,2,\cdots$, a
small axion-induced rotation at frequency $\mu$ generates sidebands at
$\omega = n\Omega_{\rm orb} \pm \mu$. For a single bright binary, we forecast
the statistical sensitivity using parameters motivated by $\mu^1$ Sco and by
current high-precision optical polarimetry. We find that the sensitivity to the
axion-photon coupling can reach
$g_{a\gamma} \sim 2.4\times 10^{-12}\,{\rm GeV}^{-1}$
at $\mu=10^{-20}\,{\rm eV}$ for a month-long observation with a cadence of order ten minutes.

We also discuss an extension to an array of close binaries, in analogy with
pulsar arrays. For binaries whose source positions are separated by distances larger than the axion coherence length, the source terms have independent phases,
whereas the Earth term is common to all targets. A multi-binary analysis can
therefore extract the common Earth term. For comparable targets, the statistical measurement
uncertainty scales as $N^{-1/2}$, where $N$ is the number of close binaries.
In addition, target specific variations of the polarization not related to the orbital motion can be statistically reduced. In an
optimistic future scenario with $N=14$ and ppm-level polarimetry, the projected
sensitivity can reach
$g_{a\gamma} \sim 1.3\times 10^{-13}\,{\rm GeV}^{-1}$
at $\mu=10^{-20}\,{\rm eV}$. These estimates should be interpreted as
statistical projections under the assumptions of known phase-locked templates and controlled polarimetric systematics.

The remainder of this paper is organized as follows. In
Sec.~\ref{sec:reflection}, we describe reflection-induced polarization in close
binaries and introduce a phase-locked polarization template. In
Sec.~\ref{sec:axion}, we summarize axion-induced polarization rotation and the
relevant coherence properties of ultralight axion DM. In
Sec.~\ref{sec:search}, we formulate the single-binary search and derive the
statistical sensitivity. In Sec.~\ref{sec:extension}, we extend the discussion
to multiple binaries and the extraction of the common Earth term. In
Sec.~\ref{sec:comparison}, we briefly compare the proposed method with other
birefringence searches. We conclude in Sec.~\ref{sec:conclusion}.

\section{Reflection polarization of close binary stars}
\label{sec:reflection}

\subsection{Single scattering toy model}

We describe the linear polarization produced by reflection in a close binary.
We first present a minimal single scattering model whose purpose is to capture the orbital
phase dependence of the polarization.  The overall normalization of the polarization depends
on the stellar atmosphere, the dilution by direct stellar light, and the detailed radiative
transfer, and will be treated phenomenologically below.

\begin{figure}[t]
    \centering
    \includegraphics[width=0.8\linewidth]{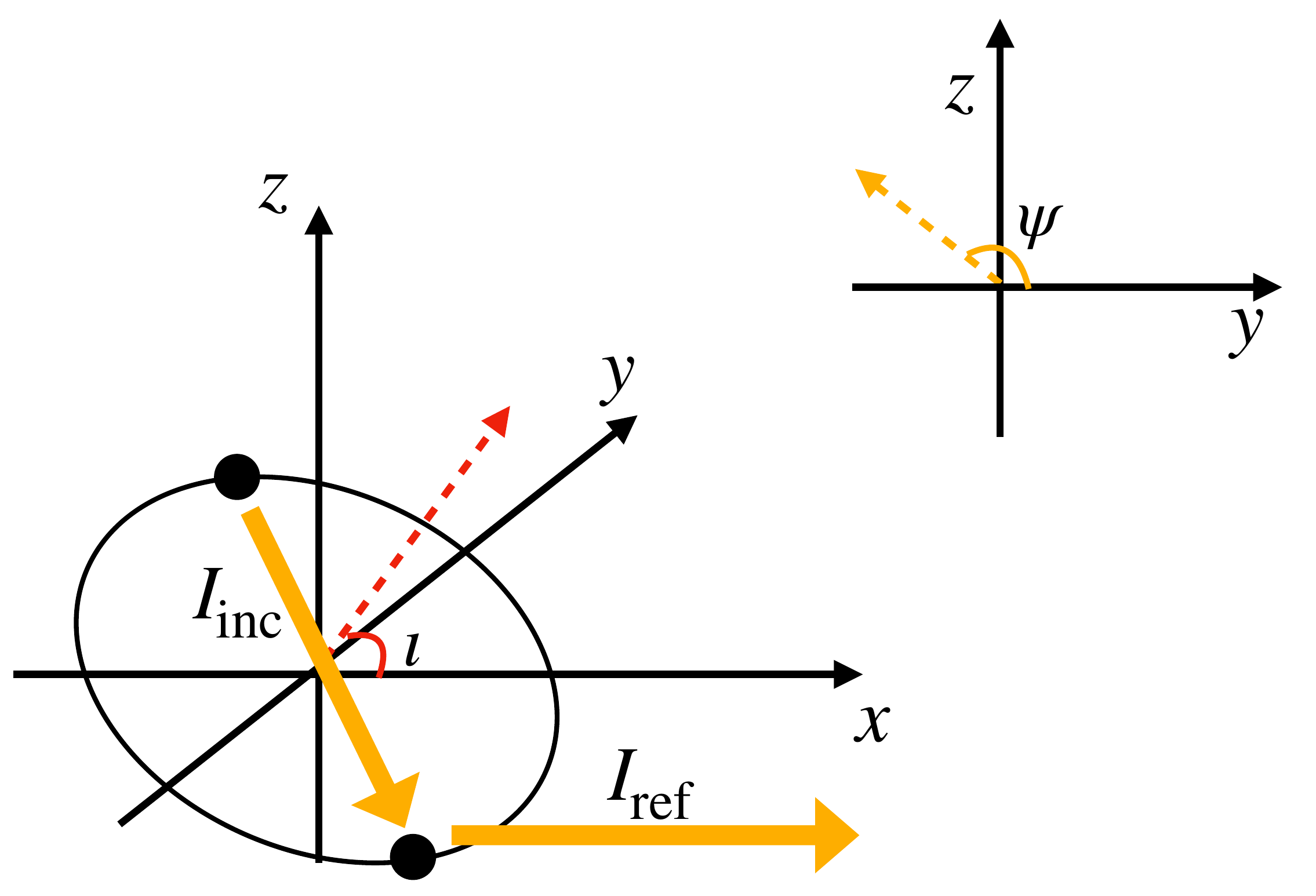}
    \caption{Geometry of the setup considered here. The observer is located along the $+\hat{\mathbf{x}}$ direction, and the polarization angle $\psi$ is measured in the $y$-$z$ plane. The binary is on a circular orbit with inclination angle $\iota$. One component of the binary illuminates the other with incident intensity $I_{\rm inc}$, and the reflected light from the illuminated star is denoted by $I_{\rm ref}$.}
    \label{fig:scat}
\end{figure}

We consider a circular orbit viewed at an inclination angle $\iota$ with respect to the line of sight (see Fig.~\ref{fig:scat}).
We place the observer along the $+\hat{\mathbf{x}}$ direction and use $(\hat{\mathbf{y}},\hat{\mathbf{z}})$ as the polarization reference axes on the sky.
For simplicity, we set the position angle of the line of nodes to $\Omega=0$.
We define the vector $\mathbf{r}(t)$ to point from the scattering component to the illuminating component:
\begin{align}
\mathbf{r}(t)
&=
a\,\hat{\mathbf{r}}(t)~,
&
\hat{\mathbf{r}}(t)
&=
(\cos\lambda \sin \iota, \sin\lambda, \cos\lambda \cos\iota)~,
\label{eq:rhat_obsframe}
\end{align}
where $a$ is the orbital separation in this toy model.
The orbital phase $\lambda$ is parameterized as
\begin{align}
\lambda(t)
&=
\Omega_{\rm orb}\,t+\lambda_0~,
&
\Omega_{\rm orb}
&=
\frac{2\pi}{T_{\rm orb}}~,
\label{eq:orb_phase_def}
\end{align}
with the orbital period $T_{\rm orb}$ and an initial phase $\lambda_0$.

Reflection-induced polarization arises when light from one star is scattered by the atmosphere of its companion and reaches the observer.
As a minimal model, we approximate the reflecting star as a point scatterer.
For a hot star, the dominant scattering mechanism is Thomson scattering by free electrons~\cite{2019NatAs...3..636B}
\footnote{For cooler stars, Rayleigh scattering by molecules becomes important. However, the angular dependence of Rayleigh scattering is the same as that of Thomson scattering~\cite{1960ratr.book.....C}, so the following discussion still applies.}.
For an unpolarized incident beam, the Stokes parameters of the reflected light in a local polarization basis whose positive-$Q$ axis is perpendicular to the scattering plane are~\cite{1960ratr.book.....C}
\begin{align}
I_{\rm ref}
&\propto
(1+\cos^2\chi)\,I_{\rm inc}~,
\cr
Q_{\rm ref}
&\propto
(1-\cos^2\chi)\,I_{\rm inc}~,
\cr
U_{\rm ref}
&=
0~,
\label{eq:stokes_scat}
\end{align}
where $I_{\rm inc}$ is the incident intensity at the scattering component.
The incident photon direction at the scatterer is $\hat{\mathbf{k}}_{\rm in}=-\hat{\mathbf{r}}$, while the outgoing direction to the observer is $\hat{\mathbf{k}}_{\rm out}=\hat{\mathbf{x}}$.
The scattering angle is therefore
\begin{align}
\cos\chi
&=
\hat{\mathbf{k}}_{\rm in}\cdot\hat{\mathbf{k}}_{\rm out}
=
-\cos\lambda\,\sin \iota~.
\label{eq:scat_angle}
\end{align}
The proportionality constant in Eq.~\eqref{eq:stokes_scat} is determined by the atmospheric structure and can in principle be computed by solving the polarized radiative-transfer equations~\cite{2019NatAs...3..636B}.

The polarization of the scattered wave is perpendicular to the scattering plane.
The observed polarization direction is aligned with the unit vector
\begin{align}
\hat{\bm{\epsilon}}(\lambda)
&=
\frac{\hat{\mathbf{k}}_{\rm out}\times \hat{\mathbf{k}}_{\rm in}}
{|\hat{\mathbf{k}}_{\rm out}\times \hat{\mathbf{k}}_{\rm in}|}~,
\label{eq:epsilon_def}
\end{align}
which lies in the $y$-$z$ plane.
The polarization angle $\psi$, defined as the angle between $\hat{\bm{\epsilon}}$ and the reference axis $\hat{\mathbf{y}}$, satisfies
\begin{align}
\cos\psi
&=
\hat{\bm{\epsilon}}\cdot \hat{\mathbf{y}}
=
\frac{\cos\lambda \cos \iota}
{\sqrt{\sin^2 \lambda + \cos^2\lambda \cos^2\iota}}~.
\label{eq:psi_def}
\end{align}
The observed Stokes parameters in the fixed sky basis are obtained by a rotation,
\begin{align}
Q_s
&=
Q_{\rm ref}\cos2\psi~,
&
U_s
&=
Q_{\rm ref}\sin2\psi~.
\label{eq:rotate_ref_stokes}
\end{align}
Using Eqs.~\eqref{eq:rhat_obsframe}--\eqref{eq:psi_def}, one finds that the phase dependence simplifies to
\begin{align}
Q_s(\lambda)
&\propto
-\frac{1}{2}\sin^2\iota
+
\frac{1}{2}(1+\cos^2 \iota)\cos 2\lambda~,
\notag\\
U_s(\lambda)
&\propto
-\cos\iota \sin2\lambda~.
\label{eq:toy_q_u}
\end{align}
Thus, in this toy model, the intrinsic polarization is phase-locked to the orbital motion.
Its time-dependent part appears entirely at the second harmonic of the orbital frequency.

The observed fractional Stokes parameters are obtained after dilution by the total observed flux,
\begin{align}
q_s
&=
\frac{Q_s}{I_{\rm tot}}~,
&
u_s
&=
\frac{U_s}{I_{\rm tot}}~.
\label{eq:fractional_stokes_binary}
\end{align}
Here $I_{\rm tot}$ includes the direct light from both stars and the reflected or scattered light.
In practice, $I_{\rm tot}$ is usually dominated by the direct stellar light. Again, the overall size of the Stokes parameters $Q_s$ and $U_s$ is determined by the atmospheric structure and detailed radiative transfer, which are not included in the present toy model.
The polarization amplitude is given as
\begin{align}
p=\sqrt{q_s^2+u_s^2},
\end{align}
which is estimated to be of order $10^{-4}$--$10^{-3}$ for the observed systems such as Spica~\citep{2019NatAs...3..636B} and $\mu^1$ Sco \citep{cotton2020phase}.

\subsection{From the toy model to realistic modeling}

Realistic systems are more complicated and involve several additional effects, but the phase locked nature of the intrinsic polarization is robust.
In the toy model above, both the illuminating source and the scattering component were approximated as pointlike objects.
In an actual binary, finite stellar radii change the visible illuminated area and can also produce eclipses.
The two stars may have different luminosities, radii, and atmospheric structures.
These effects can generate harmonic components beyond the second harmonic~\cite{1978A&A....68..415B}.
However, provided that the relevant scattering structures are stationary in the co-rotating frame of the binary, their contribution remains a periodic function of the orbital phase and hence can be expanded in harmonics of $\Omega_{\rm orb}$.

Realistic atmospheres have finite optical depth and may include absorption by species other than electrons.
Multiple scattering can also occur.
These effects can reduce the observed polarization amplitude~\cite{1978A&A....68..415B}, but they do not introduce frequencies unrelated to the orbital motion as long as the scattering structure is stationary in the co-rotating frame.

Reflection induced polarization from stellar surfaces has recently been observed.
For example, a polarization degree of $\mathcal{O}(10^{-4})$ has been measured in Spica~\cite{2019NatAs...3..636B} and $\mu^1$ Sco~\cite{cotton2020phase}.
In these systems, polarized radiative-transfer calculations reproduce the observed oscillation of  polarization.
In particular, by fitting the polarimetric data with theoretically computed templates, binary parameters such as the position angle of the line of nodes can be inferred.

In what follows, we therefore treat the intrinsic polarization of a target binary as a known periodic template.
We assume that the intrinsic polarization is phase-locked and can be written as
\begin{align}
q_s(\lambda)
&=
q_0
+
\sum_{n=1}^{N_h}
\left[
a_n e^{i n \lambda}
+
\bar{a}_n e^{-i n \lambda}
\right]~,
\notag\\
u_s(\lambda)
&=
u_0
+
\sum_{n=1}^{N_h}
\left[
b_n e^{i n \lambda}
+
\bar{b}_n e^{-i n \lambda}
\right]~.
\label{eq:template_fourier}
\end{align}
Here $q_0$ and $u_0$ are real constants, and the overbar denotes complex conjugation.
The integer $N_h$ sets the maximum harmonic included in the template; in practice, it is usually sufficient to retain only the lowest few harmonics.
This assumption is motivated by classic analyses of phase-locked binary polarization~\cite{1978A&A....68..415B,1978ApJ...221..200R}.
For the pure-reflection toy model, only the constant and $n=2$ components are present (see Eq.~\eqref{eq:toy_q_u}), while higher harmonics can account for finite-size effects, eclipses, and other departures from the idealized case. The constants $q_0$ and $u_0$ might include foreground polarization from interstellar matter. In the sensitivity estimates below, however, we only use polarization of the phase-locked component generated by the binary. 
The coefficients $a_n$ and $b_n$ can be determined either from polarized radiative-transfer modeling or by fitting long-term observational data.
In either case, we assume that the intrinsic polarization template has already been determined.

Polarization variability produced from astrophysical processes other than the orbital motion of the binary is not included in the phase-locked template in Eq.~\eqref{eq:template_fourier}. The variability that changes the amplitude is less problematic since it is orthogonal to the rotation. However, the variability that produces time-dependent rotation of the polarization may become problematic. For example, stellar pulsations cause displacements of the stellar surface with periods ranging from minutes to days depending on the pulsation mode \citep{2022ARAA..60...31K}, and may cause the variation in the structure of the atmosphere, which plays a central role in determining the polarization of the reflective photons. As seen later, the expected pulsation period may overlap the axion mass scale that we try to trace. 

Another possibility is the effect of circumstellar matter around the binary. Phase-locked polarization detection is feasible not only for detached binaries such as Spica \citep{2019NatAs...3..636B} but also for semi-detached binaries such as $\mu^1$~Sco \citep{cotton2020phase}. In fact, a semi-detached binary undergoes time-dependent mass transfer, and in a framework of non-conservative mass transfer, a fraction of the transferred gas could escape the binary and be distributed as circumstellar matter \citep{1997AA...327..620S}. Although the reflection of lights by the circumstellar matter can happen, its importance relative to phase-locked polarization from orbital motion has been less discussed yet. 

While we suppose that these astrophysical factors producing time-variable polarization other than the phase-locked template could be treated as a secondary effect as assumed in previous works \citep{2019NatAs...3..636B,cotton2020phase}, these should be included as an additional noise contribution or modeled explicitly in real data analyses. An observation of a single binary can be biased if such variability has a coherent component near the axion oscillation frequency. When applying our framework to real data, close binaries with large variability from these effects should be excluded. Still, when multiple targets are combined (Sec.~\ref{sec:extension}), variability local to each source is not expected to share a common phase across unrelated binaries and can therefore be treated as independent noise among the targets.

\section{Polarization rotation induced by axion DM}
\label{sec:axion}

We briefly review the polarization rotation induced by the axion-photon coupling.
A more detailed derivation can be found in, e.g., Refs.~\cite{Chigusa:2019,Fedderke:2019}.
We use natural units $c=\hbar=1$ and the convention
$\tilde F^{\mu\nu} \equiv \epsilon^{\mu\nu\rho\sigma}F_{\rho\sigma}/2$ with
$\epsilon^{0123}=+1$.
The relevant action is
\begin{align}
S
&=
\int d^4x
\left[
-\frac14 F_{\mu\nu}F^{\mu\nu}
+\frac12 \partial_\mu a\,\partial^\mu a
-\frac12 \mu^2 a^2\right.\cr
&\qquad\qquad\qquad \left.
-\frac{g_{a\gamma}}{4}aF_{\mu\nu}\tilde F^{\mu\nu}
\right]~.
\label{eq:action_axion}
\end{align}
In a slowly varying axion background, the two circular polarizations of light acquire
different phases.  In the geometric-optics limit, and neglecting corrections suppressed by
the dark-matter velocity, the resulting rotation angle of linear polarization is given by 
\begin{align}
\theta_a(t_{\rm obs})
=
\frac{g_{a\gamma}}{2}
\left[
a(t_{\rm obs},\mathbf{x}_{\rm obs})
-
a(t_{\rm em},\mathbf{x}_{\rm em})
\right]~.
\label{eq:theta_endpoint}
\end{align}
The overall sign of $\theta_a$ depends on the convention for circular polarization and for
the positive Stokes-$U$ direction.  Throughout this paper we adopt the convention in which
a positive $\theta_a$ acts on the Stokes parameters as
\begin{align}
\begin{pmatrix}
Q_{\rm obs} \\
U_{\rm obs}
\end{pmatrix}
=
\begin{pmatrix}
\cos 2\theta_a & -\sin 2\theta_a\\
\sin 2\theta_a & \cos 2\theta_a
\end{pmatrix}
\begin{pmatrix}
Q_{s}\\
U_{s}
\end{pmatrix}~.
\label{eq:stokes_rotation}
\end{align}
For $|\theta_a|\ll1$, this reduces to
\begin{align}
Q_{\rm obs}
&\sim
Q_s-2\theta_a U_s~,
&
U_{\rm obs}
&\sim
U_s+2\theta_a Q_s~.
\label{eq:small_angle_QU}
\end{align}

For a source at distance $D$, we write $t_{\rm em}=t-D$ and decompose the signal into an
Earth term and a source term,
\begin{align}
\theta_a(t)
&=
\theta_E(t)-\theta_S(t)~,
\label{eq:theta_ES_decomp}
\\
\theta_E(t)
&=
\frac{g_{a\gamma}}{2}a(t,\mathbf{x}_E)~,
&
\theta_S(t)
&=
\frac{g_{a\gamma}}{2}a(t-D,\mathbf{x}_S)~.
\label{eq:theta_ES_def}
\end{align}

The Galactic axion DM field has a large occupation number and can be treated as a
classical wave~\cite{Hui:2016ltb}.  At a fixed position $\mathbf{x}$, over an observing time much shorter than the coherence
time, the field is well approximated by the monochromatic wave
\begin{align}
a(t,\mathbf{x})
\sim
a_0(\mathbf{x})\cos\left[\mu t+\phi(\mathbf{x})\right]~,
\label{eq:axion_coherent}
\end{align}
with $\phi(\mathbf{x})$ denoting the phase of the field at $\mathbf{x}$. The phase is uncorrelated between the points separated by distances larger than the coherence length. 
The coherence time and coherence length can be estimated to be 
\begin{align}
\tau_c
&\sim
\frac{1}{\mu v^2}
\sim
2\times10^4\,{\rm yr}
\left(\frac{\mu}{10^{-21}\,{\rm eV}}\right)^{-1}
\left(\frac{v}{10^{-3}}\right)^{-2}~,
\label{eq:coh_time}
\\
\ell_c
&\sim
\frac{1}{\mu v}
\sim
6\,{\rm pc}
\left(\frac{\mu}{10^{-21}\,{\rm eV}}\right)^{-1}
\left(\frac{v}{10^{-3}}\right)^{-1}~.
\label{eq:coh_length}
\end{align}
Note that the amplitude $a_0$ is related to the energy density $\rho_a$ as
\begin{align}
\rho_a(\mathbf{x})
\sim
\frac12 \mu^2 a_0^2(\mathbf{x})~.
\label{eq:rho_axion}
\end{align}
Thus, the rotation angle can be written as
\begin{align}
\theta_a(t)
&=
\theta_{E,0}\cos(\mu t+\phi_E)
-
\theta_{S,0}\cos(\mu t+\phi_S)~,
\label{eq:theta_general_ES}
\\
\theta_{E,0}
&=
\frac{g_{a\gamma}}{2}
\frac{\sqrt{2\rho_E}}{\mu}~,
\qquad 
\theta_{S,0}
=
\frac{g_{a\gamma}}{2}
\frac{\sqrt{2\rho_S}}{\mu}~,
\label{eq:theta_ES_amplitudes}
\end{align}
where $\rho_E$ and $\rho_S$ denote the energy densities at the Earth and at the source, respectively. We have absorbed the propagation phase $-\mu D$ into the definition of $\phi_S$.

For the sensitivity estimates below, we take $\rho_S\sim \rho_E = \rho_a$ and define
\begin{align}
\theta_{a,0}
&=
\frac{g_{a\gamma}}{2}
\frac{\sqrt{2\rho_a}}{\mu}\cr
&= 6.1\times 10^{-3}\, {\rm deg}
\left(
\frac{g_{a\gamma}}{10^{-12}\,{\rm GeV}^{-1}}
\right)\cr
&\qquad \times 
\left(
\frac{\rho_a}{0.3\,{\rm GeV/cm^3}}
\right)^{1/2}
\left(
\frac{\mu}{10^{-20}\,{\rm eV}}
\right)^{-1}~.
\label{eq:theta_a0_def}
\end{align}
Then
\begin{align}
\theta_a(t)
=
\theta_{a,0}
\left[
\cos(\mu t+\phi_E)
-
\cos(\mu t+\phi_S)
\right]~.
\label{eq:theta_equal_density}
\end{align}

For binaries with $D\gg\ell_c$, the source phase is effectively uncorrelated with the
Earth phase.  Source phases of different binaries are also independent if their separations
are larger than $\ell_c$.  In contrast, the Earth term is common to all targets observed
within the coherence time.  This distinction will be used in Sec.~\ref{sec:extension} to
separate the common Earth term from target-dependent source terms, in analogy with
pulsar polarization arrays~\cite{Liu:2021zlt}.

\section{Searching for axion DM with binary polarimetry}
\label{sec:search}

In this section, we formulate an axion search using the phase-locked polarization of a
close binary. The basic idea is the same as in other birefringence searches using linearly
polarized sources. If the intrinsic polarization is known, an axion-induced rotation can be searched for as a small time-dependent rotation relative to the intrinsic polarization template.
The distinctive feature of close binaries is that the intrinsic polarization itself is periodic
and contains harmonics of the orbital frequency. However, we show that the time dependence of the template does not produce any difficulty.

\subsection{Axion-induced modulation of the reflection polarization}
\label{sec:single}

Let $(q_s(t),u_s(t))$ denote the intrinsic fractional Stokes parameters of the binary.  In the presence of an axion-induced rotation angle $\theta_a(t)$, the observed Stokes parameters are
\begin{align}
\begin{pmatrix}
q_{\rm obs}(t)\\
u_{\rm obs}(t)
\end{pmatrix}
=
\begin{pmatrix}
\cos 2\theta_a(t) & -\sin 2\theta_a(t)\\
\sin 2\theta_a(t) & \cos 2\theta_a(t)
\end{pmatrix}
\begin{pmatrix}
q_s(t)\\
u_s(t)
\end{pmatrix}~.
\label{eq:qu_rotation_single}
\end{align}
For $|\theta_a|\ll1$,
\begin{align}
q_{\rm obs}(t)
&\sim
q_s(t)-2\theta_a(t)u_s(t)~,\cr
u_{\rm obs}(t)
&\sim
u_s(t)+2\theta_a(t)q_s(t)~.
\label{eq:qu_small_rotation_single}
\end{align}
Introducing the complex Stokes parameter $z(t)= q(t)+iu(t)$, we can write the observed complex Stokes parameter as
\begin{align}
z_{\rm obs}(t)
\sim
z_s(t)+2i\theta_a(t)z_s(t)~.
\label{eq:z_small_rotation}
\end{align}
Using the Fourier template in Eq.~\eqref{eq:template_fourier}, and absorbing the initial
orbital phase $\lambda_0$ into the complex coefficients, we write
\begin{align}
z_s(t)
&=
z_0
+
\sum_{n=1}^{N_h}
\left[
C_n e^{in\Omega_{\rm orb}t}
+
D_n e^{-in\Omega_{\rm orb}t}
\right]~,
\label{eq:z_template_fourier}
\\
z_0
&=
q_0+i u_0~,\\
C_n
&=
(a_n+i b_n)e^{in\lambda_0}~,\\
D_n
&=
(\bar a_n+i\bar b_n)e^{-in\lambda_0}~.
\label{eq:complex_template_coeffs}
\end{align}

Substituting Eq.~\eqref{eq:theta_equal_density} into Eq.~\eqref{eq:z_small_rotation}, we obtain
\begin{align}
&z_{\rm obs}(t)-z_s(t)
= 2 i \theta_{a}(t) z_0 \cr
&+
\theta_{a,0}
\sum_{n=1}^{N_h}
\biggl[
c_{n,+}e^{i(n\Omega_{\rm orb}+\mu)t}
+
c_{n,-}e^{i(n\Omega_{\rm orb}-\mu)t}
\cr
&
+
d_{n,+}e^{-i(n\Omega_{\rm orb}+\mu)t}
+
d_{n,-}e^{-i(n\Omega_{\rm orb}-\mu)t}
\biggr]~,
\label{eq:sideband_expansion}
\end{align}
where
\begin{align}
c_{n,\pm}
&=
i\left(e^{\pm i\phi_E}-e^{\pm i\phi_S}\right)C_n~,
\cr
d_{n,\pm}
&=
i\left(e^{\mp i\phi_E}-e^{\mp i\phi_S}\right)D_n~.
\label{eq:sideband_d_coeffs}
\end{align}
Thus, the axion signal appears as sidebands around the orbital harmonics of the intrinsic polarization template. Furthermore, the ratio of each sideband coefficient to the corresponding intrinsic Fourier coefficient is independent of $n$. This sideband structure helps separate the intrinsic polarization template from the axion signal. 

When the axion oscillation frequency is close to an integer multiple of the orbital frequency, the sidebands can overlap with intrinsic orbital harmonics. This situation can occur in the present setting, since the orbital period is of order a day, for which the corresponding axion mass is around $10^{-20}{\rm eV}$. For such axion masses, the axion signal becomes partially degenerate with the intrinsic polarization template, which can degrade the sensitivity. 
We neglect such degeneracies in the estimates below for simplicity.

Given the template, an estimator for the instantaneous rotation angle is
\begin{align}
\hat\theta_a(t)
=
\frac{
q_s(t)u_{\rm obs}(t)-u_s(t)q_{\rm obs}(t)
}{
2P_s^2(t)
}~,
\label{eq:theta_estimator}
\end{align}
where
\begin{align}
P_s(t)
\equiv
\sqrt{q_s^2(t)+u_s^2(t)}
\label{eq:Ps_def}
\end{align}
is the polarization amplitude of the template.  The measured Stokes parameters are given by the sum of the signal and noise,
\begin{align}
\begin{pmatrix}
q_{\rm obs}(t)\\
u_{\rm obs}(t)
\end{pmatrix}
=
\begin{pmatrix}
\cos 2\theta_a(t) & -\sin 2\theta_a(t)\\
\sin 2\theta_a(t) & \cos 2\theta_a(t)
\end{pmatrix}
\begin{pmatrix}
q_s(t)\\
u_s(t)
\end{pmatrix}
+
\begin{pmatrix}
n_q(t)\\
n_u(t)
\end{pmatrix}~.
\label{eq:qu_with_noise}
\end{align}
In the following, we assume that the noise is Gaussian with zero mean and satisfies
\begin{align}
\langle n_q^2\rangle
&=
\langle n_u^2\rangle
=
\sigma_p^2~,
&
\langle n_q n_u\rangle
&=
0~,
\label{eq:noise_assumption}
\end{align}
Here, the angle brackets denote ensemble averages. The parameter $\sigma_p$ represents the polarimetric measurement uncertainty and may also include  stochastic variability of the polarization local to the source. 
The variance of the estimator \eqref{eq:theta_estimator} is then
\begin{align}
\sigma_\theta^2(t)
\sim
\frac{\sigma_p^2}{4P_s^2(t)}~.
\label{eq:theta_variance}
\end{align}

\subsection{Sensitivity to the axion-photon coupling}

We now estimate the sensitivity to the axion-photon coupling for a single binary observed over a total time $T_{\rm obs}$
with cadence $t_{\rm cad}$.  We assume $T_{\rm obs}\ll\tau_c$, so that the axion signal is
monochromatic during the observation, and $D\gg\ell_c$, so that the source phase is
independent of the Earth phase.  The rotation angle can then be modeled as
\begin{align}
\theta_a(t)
=
A\cos\mu t+B\sin\mu t~,
\label{eq:theta_AB_model}
\end{align}
where
\begin{align}
A
&=
\theta_{a,0}\left(\cos\phi_E-\cos\phi_S\right)~,
&
B
&=
-\theta_{a,0}\left(\sin\phi_E-\sin\phi_S\right)~.
\label{eq:AB_source_earth}
\end{align}

Let $t_k$ be the observation times, with $k=1,\cdots,M$ and
$M=T_{\rm obs}/t_{\rm cad}$. We define $d_k$ as the observed value of $\hat{\theta}_{a}(t)$ at $t = t_k$, and $\sigma_k^2$ as its variance, 
\begin{align}
d_k
&=
\hat\theta_a(t_k)~,
&
\sigma_k^2
&=
\sigma_\theta^2(t_k)~.
\label{eq:data_def_single}
\end{align}
The log-likelihood for the parameters $\mathbf{a}=(A,B)^T$ at fixed $\mu$ is
\begin{align}
\ln{\cal L}(\mathbf{a}|\mu)
&=
-\frac12
\sum_{k=1}^{M}
\frac{
\left[
d_k-A\cos(\mu t_k)-B\sin(\mu t_k)
\right]^2
}{
\sigma_k^2
}
+{\rm const.}
\nonumber\\
&=
-\frac12
(\mathbf{d}-X\mathbf{a})^T
\Sigma^{-1}
(\mathbf{d}-X\mathbf{a})
+{\rm const.}~,
\label{eq:loglike_single}
\end{align}
where $\mathbf{d} = (d_1, \dots , d_M)^T$, and the components of the matrix $X$ are
\begin{align}
X_{k1}
&=
\cos(\mu t_k)~,
&
X_{k2}
&=
\sin(\mu t_k)~,
\label{eq:design_matrix_single}
\end{align}
The noise covariance matrix $\Sigma$ is assumed to be diagonal,
\begin{align}
\Sigma
&=
{\rm diag}(\sigma_1^2,\cdots,\sigma_M^2)~.
\label{eq:design_matrix_single-2}
\end{align}

Since the model is linear in $A$ and $B$, the maximum-likelihood estimator and its
covariance matrix are
\begin{align}
\hat{\mathbf{a}}
&=
(X^T\Sigma^{-1}X)^{-1}X^T\Sigma^{-1}\mathbf{d}~,
&
{\rm Cov}(\hat{\mathbf{a}})
&=
F^{-1}~,
\label{eq:linear_estimator_single}
\\
F
&\equiv
X^T\Sigma^{-1}X~.
\label{eq:fisher_single}
\end{align}
For data spanning many axion oscillation periods with sufficient sampling,
$T_{\rm obs}\gg\mu^{-1}$, the off-diagonal element of $F$ is small, and the uncertainties in the coefficients $A$ and $B$ are given by
\begin{align}
\sigma(A)
\sim
\sigma(B)
\sim
\left(
\frac12\sum_{k=1}^{M}\frac{1}{\sigma_k^2}
\right)^{-1/2}~.
\label{eq:sigma_A_B_single}
\end{align}

The fitted coefficients $(A,B)$ measure the difference between the Earth and source
terms. The amplitude of the coefficients is
\begin{align}
\sqrt{A^2+B^2}
=
2\theta_{a,0}
\left|
\sin\frac{\phi_E-\phi_S}{2}
\right|~.
\label{eq:endpoint_amplitude_single}
\end{align}
For the sensitivity estimates below, we replace
$\left|\sin[(\phi_E-\phi_S)/2]\right|$ by its root mean square value $1/\sqrt2$.  This gives
\begin{align}
\sigma(\theta_{a,0})
\sim
\left(
\sum_{k=1}^{M}\frac{1}{\sigma_k^2}
\right)^{-1/2}~.
\label{eq:theta0_sensitivity_general_single}
\end{align}
Substituting Eq.~\eqref{eq:theta_variance}, we obtain
\begin{align}
\sigma(\theta_{a,0})
&\sim
\frac{\sigma_p}{2P_{\rm rms}}
\left(
\frac{t_{\rm cad}}{T_{\rm obs}}
\right)^{1/2}~,
\label{eq:theta0_sensitivity_single}
\\
P_{\rm rms}^2
&=
\frac{1}{M}
\sum_{k=1}^{M}P_s^2(t_k)~.
\label{eq:Prms_def}
\end{align}
Using Eq.~\eqref{eq:theta_a0_def}, the corresponding uncertainty in the axion photon-coupling is
\begin{align}
\sigma(g_{a\gamma})
&\sim
\frac{\sqrt2\,\mu}{\sqrt{\rho_a}}\,
\sigma(\theta_{a,0})
\nonumber\\
&\sim
2.4\times10^{-12}\,{\rm GeV}^{-1}
\left(\frac{\mu}{10^{-20}\,{\rm eV}}\right)\cr
&\qquad 
\times \left(\frac{\rho_a}{0.3\,{\rm GeV/cm^3}}\right)^{-1/2}
\left(\frac{300\,{\rm ppm}}{P_{\rm rms}}\right)
\left(\frac{\sigma_p}{10\,{\rm ppm}}\right)
\cr
&\qquad \qquad 
\times
\left(\frac{t_{\rm cad}}{10\,{\rm min}}\right)^{1/2}
\left(\frac{T_{\rm obs}}{30\,{\rm day}}\right)^{-1/2}~.
\label{eq:single_g_sensitivity}
\end{align}
The benchmark value $\sigma_p\sim10\,{\rm ppm}$ is motivated by high-precision optical
polarimeters such as HIPPI-2~\cite{Bailey:2020HIPPI2}.  The values
$P_{\rm rms}\sim 300\,{\rm ppm}$ and $t_{\rm cad}\sim10\,{\rm min}$ are motivated by the
observed phase-locked polarization of $\mu^1$ Sco~\cite{cotton2020phase}. In terms of the polarization angle, the uncertainty corresponds to $\sigma(\theta) \sim 1.5 \times 10^{-2}\, {\rm deg}$. We assume that any systematic error in the polarization angle that coherently
oscillates on hour-to-month time scales is reduced below this level. If the systematic error is larger, then the sensitivity is degraded. For example, an error of $0.1\,{\rm deg}$ worsen the sensitivity by about one order of magnitude.

The accessible axion mass range is set by the cadence and the observation baseline.  The
upper edge is approximately the Nyquist frequency,
\begin{align}
\mu_{\rm max}
\sim
\frac{\pi}{t_{\rm cad}}
\sim
3.4\times10^{-18}\,{\rm eV}
\left(\frac{10\,{\rm min}}{t_{\rm cad}}\right)~,
\label{eq:mu_max}
\end{align}
while the lower edge is set by the longest oscillation period that can be resolved,
\begin{align}
\mu_{\rm min}
\sim
\frac{2\pi}{T_{\rm obs}}
\sim
1.6\times10^{-21}\,{\rm eV}
\left(\frac{30\,{\rm day}}{T_{\rm obs}}\right)~.
\label{eq:mu_min}
\end{align}
With a cadence of order ten minutes and an observing baseline of order one month, the
method is naturally suited to axion periods from hours to days, corresponding roughly to
$\mu\sim10^{-20}$--$10^{-18}\,{\rm eV}$.

\section{Extension to many binaries}
\label{sec:extension}

So far, we have considered an axion search using a single bright binary.  A single target measures only the difference between the axion fields at the Earth and at the source, and the source term has an unknown phase. If multiple binaries are observed, the source terms have independent phases for targets separated by distances larger than the axion coherence length, whereas the Earth term is common to all targets.
Thus, an array of binaries can be used to extract the common Earth term, in analogy with pulsar polarization arrays~\cite{Liu:2021zlt}.

The analysis in Sec.~\ref{sec:single} can be extended to an ensemble of binaries. For simplicity, we assume that all targets are observed for the same duration and with the same cadence, and that they have comparable values of the polarization amplitude $P_{\rm rms}$. The observations of different targets need not be simultaneous, as long as the total observing span is shorter than the axion coherence time. For each binary $a=1,\cdots,N$, we first estimate the coefficients
$$
(A^{(a)},B^{(a)})
$$
using the likelihood in Eq.~\eqref{eq:loglike_single}.  The Fisher matrix for target $a$ is
\begin{align}
F^{(a)}
=
(X^{(a)})^T
(\Sigma^{(a)})^{-1}
X^{(a)}~,
\label{eq:fisher_many}
\end{align}
where the matrices $X^{(a)}$ and $\Sigma^{(a)}$ are defined as in Eqs.~\eqref{eq:design_matrix_single} and \eqref{eq:design_matrix_single-2}, but evaluated at the observation times of that target.

The fitted coefficients can be written as
\begin{align}
A^{(a)}
&=
A_E-A_S^{(a)}~,
&
B^{(a)}
&=
B_E-B_S^{(a)}~,
\label{eq:AB_many_decomp}
\end{align}
where
\begin{align}
A_E
&=
\theta_{a,0}\cos\phi_E~,
&
B_E
&=
-\theta_{a,0}\sin\phi_E~,
\label{eq:Earth_coeffs}
\\
A_S^{(a)}
&=
\theta_{a,0}\cos\phi_{S,a}~,
&
B_S^{(a)}
&=
-\theta_{a,0}\sin\phi_{S,a}~,
\label{eq:source_coeffs}
\end{align}
with $\phi_{S,a}$ denoting the source phase for target $a$.
For well-sampled data, the uncertainty for target $a$ is
\begin{align}
s_a
=
\sigma(A^{(a)})
\sim
\sigma(B^{(a)})
\sim
\left(
\frac12
\sum_{k}
\frac{1}{(\sigma_k^{(a)})^2}
\right)^{-1/2}~,
\label{eq:sa_def}
\end{align}
where $(\sigma^{(a)}_k)^2$ denotes the variance of the estimator $\hat{\theta}_a$ for target $a$ at $t = t_k$, defined as in Eq.~\eqref{eq:data_def_single}.

We estimate the common Earth term coefficients using the weighted averages
\begin{align}
\hat A_E
&=
\frac{\sum_{a=1}^{N} A^{(a)}/s_a^2}{\sum_{a=1}^{N}1/s_a^2}~,
&
\hat B_E
&=
\frac{\sum_{a=1}^{N} B^{(a)}/s_a^2}{\sum_{a=1}^{N}1/s_a^2}~.
\label{eq:weighted_Earth_estimator}
\end{align}
Since the source phases $\phi_{S,a}$ are random among the sources, only the contribution of the Earth term remains after averaging. However, for a finite number of sources, the random source phases contribute to the variance. The variance of the Earth term estimator is calculated to be
\begin{align}
\sigma^2(\hat A_E)
=
\sigma^2(\hat B_E)
\sim
\left(
\sum_{a=1}^{N}\frac{1}{s_a^2}
\right)^{-1}\left(\frac{\theta_{a,0}^2}{2}\frac{\sum_a 1/s_a^4}{\sum_a 1/s_a^2} + 1\right)~.
\label{eq:measurement_variance_many}
\end{align}
Assuming that all targets have comparable uncertainty, $s_a=s$, and using Eqs.~\eqref{eq:sa_def} and~\eqref{eq:theta_variance}, Eq.~\eqref{eq:measurement_variance_many} becomes
\begin{align}
\sigma(\hat{A}_E)
=
\sigma(\hat{B}_E)
\sim
\frac{\sigma_p}{P_{\rm rms}}
\left(
\frac{t_{\rm cad}}{2 N T_{\rm obs}}
\right)^{1/2}~.
\label{eq:Earth_component_uncertainty_explicit}
\end{align}
Here, we neglect the source-phase contribution, since it modifies the following result only by $\mathcal{O}(1/N)$. 
Since $A_E^2+B_E^2=\theta_{a,0}^2$,
the uncertainty in the rotation angle is
\begin{align}
\sigma(\theta_{a,0})_{\rm multi}
\sim
\frac{\sigma_p}{P_{\rm rms}}
\left(
\frac{t_{\rm cad}}{2N T_{\rm obs}}
\right)^{1/2}~.
\label{eq:theta0_many}
\end{align}
Thus, the uncertainty scales as $N^{-1/2}$ for the common Earth term. The corresponding sensitivity to the axion-photon coupling is
\begin{align}
\sigma(g_{a\gamma})_{\rm multi}
&\sim
\frac{\sqrt2\,\mu}{\sqrt{\rho_a}}\,
\sigma(\theta_{a,0})_{\rm multi}
\nonumber\\
&\sim
1.3\times10^{-13}\,{\rm GeV}^{-1}
\left(\frac{\mu}{10^{-20}\,{\rm eV}}\right)\cr
&\qquad 
\times \left(\frac{\rho_a}{0.3\,{\rm GeV/cm^3}}\right)^{-1/2}
\left(\frac{\sigma_p}{1\,{\rm ppm}}\right)
\left(\frac{200\,{\rm ppm}}{P_{\rm rms}}\right)
\cr
&\qquad
\times
\left(\frac{t_{\rm cad}}{10\,{\rm min}}\right)^{1/2}
\left(\frac{T_{\rm obs}}{30\,{\rm day}}\right)^{-1/2}
\left(\frac{N}{14}\right)^{-1/2}~.
\label{eq:many_g_sensitivity}
\end{align}

To detect reflection-induced polarization with optical polarimetry, the targets must be
bright and nearby.  Since the polarization of hot binaries is mainly produced by Thomson
scattering off ionized electrons, we also require a sufficiently hot stellar atmosphere.  In App.~\ref{app:A}, we survey potentially useful binaries. In the sensitivity estimate, we use $N=14$ as an optimistic future benchmark.  We also take
$P_{\rm rms}=200\,{\rm ppm}$ as a conservative fiducial polarization amplitude motivated by
Spica.

Combining many binaries also suppresses uncorrelated stellar variability local to each target.
However, it does not suppress systematic effects that are common to the array, such as instrumental polarization angle drift or calibration errors. Such effects can mimic the Earth term signal and set a systematic floor to the sensitivity. In the present study, we ignore such errors for simplicity, but they must be accounted for in applications to real data. Any spurious common time-dependent rotation must be smaller than $\sigma(\theta_{a,0})$. For the benchmark parameters, this corresponds to $\sim 8.2 \times 10^{-4}{\rm deg}$.

\begin{figure}[t]
\centering
\includegraphics[width=1.0\linewidth]{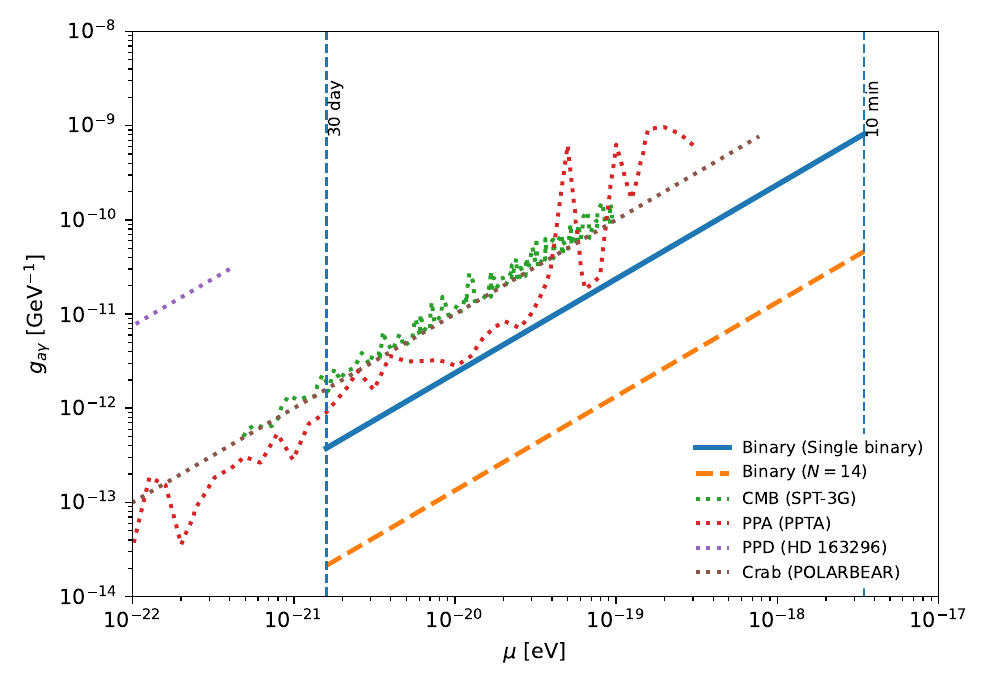}
\caption{Representative existing constraints on the axion-photon coupling $g_{a\gamma}$ from
oscillating-birefringence searches, compared with the projected 1$\sigma$ statistical sensitivity of
close binary optical polarimetry.  The blue solid and orange dashed lines show the
single-binary and multi-binary benchmarks estimated in Eqs.~\eqref{eq:single_g_sensitivity}
and~\eqref{eq:many_g_sensitivity}, respectively.  The comparison curves are shown only as
representative current birefringence constraints; the proposed binary-polarimetry method
targets a complementary high-cadence optical window.
The CMB and Pulsar Polarization Array (PPA) curves are taken from the AxionLimits compilation~\cite{OHare:2020AxionLimits}, based on the cited analyses, and are shown here for comparison. The protoplanetary disk (PPD) line is taken from~\cite{Narita:2026dvw}. The Crab nebula (Crab) is taken from~\cite{POLARBEAR:2025djl}. }
\label{fig:mu_g_constraints}
\end{figure}

\section{Relation to existing searches}
\label{sec:comparison}

The method proposed in this paper is conceptually similar to other searches for ultralight axion DM. The methods differ mainly in the polarized source used to define the intrinsic, unrotated polarization and in the time window over which the signal can be monitored. Close binaries provide a phase-locked oscillating polarization template, whereas other methods such as CMB polarization~\cite{Ferguson:2022,BICEP:2021}, protoplanetary disk polarimetry~\cite{Fujita:2018zaj,Davydov:2023hgw, Narita:2026dvw}, Crab Nebula~\cite{POLARBEAR:2025djl, POLARBEAR:2025djl}, and pulsar polarization arrays~\cite{Liu:2021zlt, Castillo:2022, EPTA:2024gxu,Yuwen:2026zjk} use different template structures and are subject to different dominant systematics. 

Figure~\ref{fig:mu_g_constraints} shows the projected statistical sensitivity of close-binary polarimetry together with representative existing birefringence constraints.
Cosmic microwave background searches~\cite{Fedderke:2019,Ferguson:2022,BICEP:2021} benefit from the large number of polarization modes, but they rely on diffuse sky polarization and are not naturally optimized for dedicated high-cadence monitoring at periods of $10$ to $10^2$ minutes.
Time domain polarimetry of a bright polarized source has also been demonstrated with the Crab Nebula.
This is the closest existing single source comparison, but it uses the polarization angle of an extended stationary source in the millimeter band, whereas the close binary method uses nonzero orbital harmonics of an optical phase locked polarization.
Searches based on protoplanetary disk polarimetry~\cite{Fujita:2018zaj,Narita:2026dvw} are closer in spirit to the template-based aspect of our proposal, since they also rely on an astrophysical scattering template.
However, their template is a spatial polarization pattern, while the close-binary template is a repeated time-domain signal.
Pulsar polarization arrays~\cite{Liu:2021zlt,Castillo:2022,EPTA:2024gxu,Yuwen:2026zjk} are the closest analog of the multi-binary strategy proposed here, because they also statistically isolate the common Earth term by combining many sources.
In contrast to radio pulsar polarimetry, optical binary polarimetry is essentially unaffected by Faraday rotation, but it requires accurate modeling or empirical reconstruction of the intrinsic binary polarization.

The main difference of close binaries from other proposed polarized sources is the possibility of high cadence monitoring.  This makes the method naturally suited to axion periods from hours to days, corresponding roughly to $\mu\sim 10^{-20}$--$10^{-18}\,{\rm eV}$. The price is that the intrinsic stellar polarization template, stellar variability, and polarization-angle calibration must be controlled.  Thus, close-binary polarimetry can in principle extend to probe higher mass range.

\section{Conclusion}
\label{sec:conclusion}

We have proposed close binary polarimetry as a probe of birefringence induced by ultralight axion DM. In close binary systems, reflection in a stellar atmosphere can generate linearly polarized light whose time dependence is locked to the orbital phase. Thus, the phase-locked polarization can provide a template for searching for a small birefringence signal. The intrinsic Stokes parameters contain harmonics of the orbital frequency, $n \Omega_{\rm orb}$ with $n = 1, 2, \cdots$, and an axion-induced polarization rotation modulates this template. In particular, an axion with mass $\mu$ generates sidebands at $n\Omega_{\rm orb} \pm \mu$. The amplitude of the sidebands is determined by the difference between the axion field values at the Earth and at the source.

For a single binary, we find that the sensitivity to the axion-photon coupling could reach
$
\sigma(g_{a\gamma})\sim 2.4\times10^{-12}\,{\rm GeV}^{-1}
$
at $\mu=10^{-20}\,{\rm eV}$. This estimate uses parameters motivated by existing observations of $\mu^1$ Sco and current high-precision optical polarimetry: polarimetric error $\sigma_p=10\,{\rm ppm}$, polarization degree $P_{\rm rms}=300\,{\rm ppm}$, cadence $t_{\rm cad}=10\,{\rm min}$, and total observation time $T_{\rm obs}=30\,{\rm days}$. The accessible mass range is set by the observing baseline and cadence, $\mu_{\rm min}\sim 2\pi/T_{\rm obs}$ and $\mu_{\rm max}\sim \pi/t_{\rm cad}$, corresponding roughly to axion masses $10^{-21}$--$10^{-18}{\rm eV}$.

We have also discussed an extension to an array of close binaries. By combining multiple binaries, one can extract the common axion signal and improve the sensitivity.  In an optimistic future scenario with $N=14$ binaries, $\sigma_p=1\,{\rm ppm}$, $P_{\rm rms}=200\,{\rm ppm}$, $t_{\rm cad}=10\,{\rm min}$, and $T_{\rm obs}=30\,{\rm days}$, the projected statistical sensitivity reaches $\sigma(g_{a\gamma})\sim 1.3\times10^{-13}\,{\rm GeV}^{-1}$ at $\mu=10^{-20}\,{\rm eV}$.  This estimate should be interpreted as a statistical projection rather than an exclusion limit, since no real polarimetric data have been analyzed in this work.

Several ingredients are required before the present method can be applied to real observations. The intrinsic binary polarization template must be reconstructed with sufficient accuracy, either from polarized radiative transfer or from long term monitoring of the binary. Furthermore, stellar variability unrelated to the orbital motion, such as pulsations, winds, or surface inhomogeneities, can generate coherent polarization signals and must be modeled appropriately. In addition, polarization-angle calibration and instrumental polarization must be taken into account. These systematics are not captured by the estimates in the present study and should be addressed in a dedicated analysis of real data.

\section*{Acknowledgement}
We thank Naoki Seto for the illuminating discussion, especially regarding the binary polarization.
T.M. was supported by JSPS KAKENHI Grant Number JP26KJ0055.
K.N. was supported by JSPS KAKENHI Grant Numbers JP24KJ0117 and JP25K17389. H. O. was supported by JSPS KAKENHI Grant Numbers JP23H00110 and JP25K17388.

\appendix

\section{Preselection of close binary targets}
\label{app:A}

In this appendix, we examine whether the known population of close binaries contains a sufficient number of targets that can be used for the axion search. Here, we do not attempt to predict the reflection-induced polarization amplitude for each cataloged system, since such a prediction requires detailed polarized radiative-transfer calculations and binary geometry. Instead, we construct a conservative catalog-level preselection of systems for which a detectable reflection signal is physically plausible. The following discussion should be regarded as a rough criterion and the final suitability of a target must be established through dedicated modeling or polarimetric observations.

The first practical requirement is that the normalized Stokes parameters can be measured with sufficiently small statistical uncertainties. The optimistic multi-binary benchmark in Sec.~\ref{sec:extension} assumes a per-epoch uncertainty $\sigma_p=1\,{\rm ppm}$ in each Stokes parameter. The photon shot-noise contribution scales as
\begin{align}
    \sigma_p \sim \frac{1}{\sqrt{N_\gamma}} \propto D^{-1} t_{\rm int}^{-1/2} 10^{0.2 m_B}~,
\end{align}
where $N_{\gamma}$ is the observed photon number, $D$ is the size of the telescope, $t_{\rm int}$ is the integration time, and $m_B$ is the apparent $B$-band magnitude of the star. From Fig.~12 of Ref.~\cite{Bailey:2020HIPPI2}, $10 {\rm ppm}$ precision is achieved for $D = 3.9\,{\rm m}, t_{\rm cad} = 10^3\,{\rm s}$ and $m = 7$. We take the limiting apparent magnitude of the binary to be the value for which $1 {\rm ppm}$ can be reached with $D = 30 {\rm m}$ and $t_{\rm cad} = 600 {\rm s}$, which gives $m_B \sim 5.9$. We therefore use $m_B \lesssim 5.9$ as an optimistic photon statistics cut. This estimate assumes a throughput comparable to HIPPI-2 after scaling the telescope to 30-meter and instrumental and calibration systematics below the ppm level, which should not be interpreted as a demonstrated total precision for a 30-meter telescope. Systems fainter than this limit may still contribute to a weighted multi-binary analysis with longer integrations or smaller statistical weight.

The second requirement is that stars in a binary should be hot enough for hydrogen atoms in the atmosphere to be ionized, similar to those in which phase-locked reflection polarization has already been observed. For the selection in the present study, we require the effective temperature $T_{\rm eff} \gtrsim 10^4{\rm K}$, according to Saha equation. From Fig.~2 of~\cite{2019NatAs...3..636B}, stars with these criteria should have a polarization degree comparable to that of Spica.

Furthermore, we require that a binary system should orbit in a short orbital period to obtain a larger reflected-light fraction. At fixed stellar radii and total mass, the reflected-light fraction scales as
\begin{align}
    \frac{I_{\rm ref}}{I_{\rm tot}}
    \propto
    \left(\frac{R_*}{a_{\rm orb}}\right)^2
    \propto P_{\rm orb}^{-4/3}~,
\end{align}
where $R_*$ is the stellar radii, $a_{\rm orb}$ is the orbital radius, and $P_{\rm orb}$ is the orbital period.
The actual polarization amplitude also depends on the luminosity ratio of the stars, atmospheric opacity, and dilution by direct stellar light. We therefore report a core sample with $P_{\rm orb}\leq5\,{\rm d}$. Since our benchmark binary Spica has a polarization degree of $200 {\rm ppm}$ with $T_{\rm eff} \sim 2 \times 10^{4}\,{\rm K}$, $\log g \sim 3.7$ and an orbital period of $4\,{\rm days}$, systems satisfying these criteria are expected to be favorable targets for reflection-induced polarization.

\begin{figure}[t]
    \centering
    \includegraphics[width=1.0\linewidth]{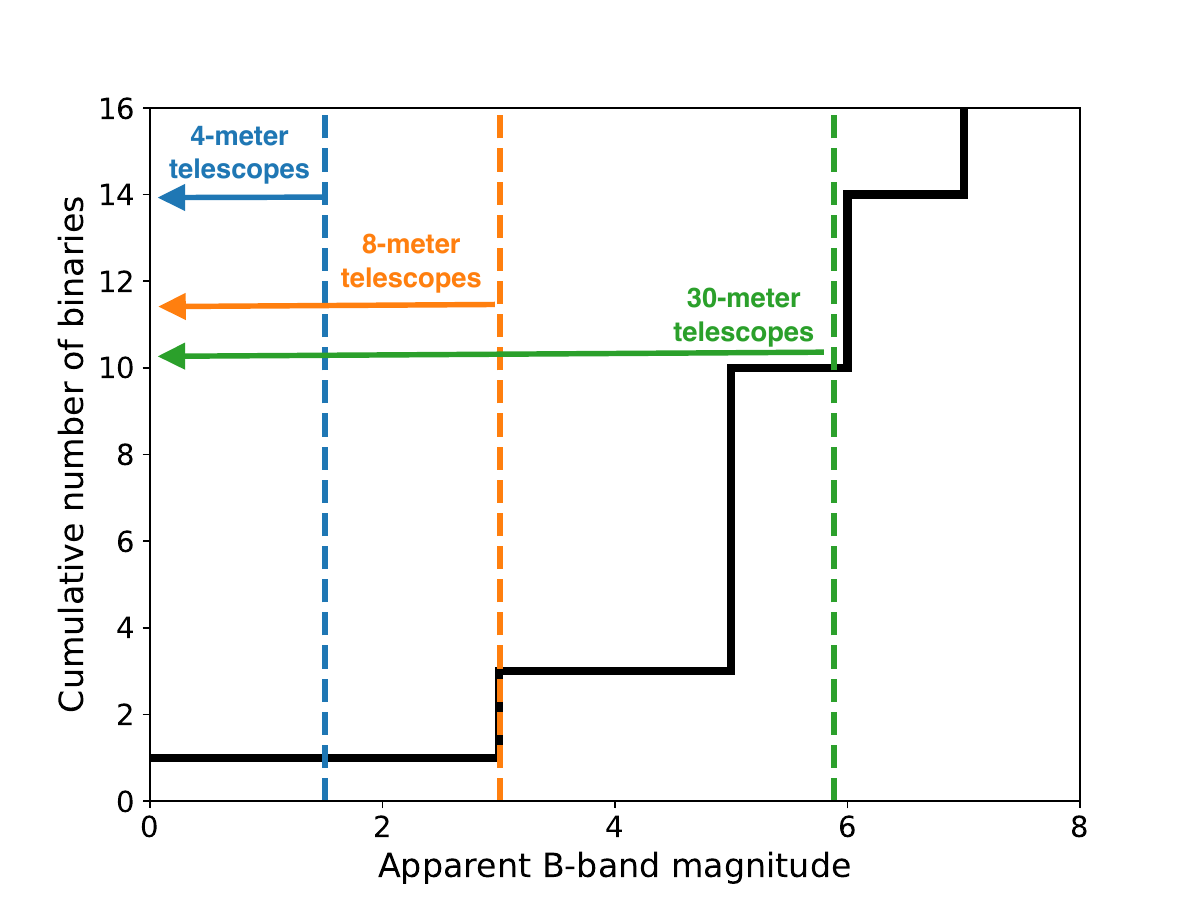}
    \caption{Cumulative number of close binaries satisfying the criteria $T_{\rm eff} \ge 10^4 {\rm K}$ and $P_{\rm orb} \le 5 {\rm days}$. We use the DEBCat catalog~\cite{2015ASPC..496..164S} and add Spica and $\mu^1$ Sco. The blue, orange, and green vertical lines show the limits on the $B$-band magnitude assuming for 4, 8, and 30 m telescopes, respectively, assuming $\sigma_p = 1 {\rm ppm}$. }
    \label{fig:cumnumber}
\end{figure}

To identify favorable targets, we survey the DEBCat catalog~\cite{2015ASPC..496..164S}, which lists eclipsing binaries with precise orbital elements. In Fig.~\ref{fig:cumnumber}, the cumulative number of candidates with $T_{\rm eff} \ge 10^4 {\rm K}$ and $P_{\rm orb} \le 5 {\rm days}$ is shown as a function of apparent B-band magnitude. From the plot, we may choose $N = 14$ as an optimistic benchmark number of future targets. That said, DEBCat is a catalog that exclusively summarizes eclipsing binaries. If we consider that non-eclipsing detached binaries can also serve as feasible targets, the number of binary targets can increase from $N=14$ in our estimate.

Note that stellar variability must be small enough that the observed polarization variation is dominated by reflection-induced polarization. As we have mentioned in Sec.~\ref{sec:reflection}, polarization variability unrelated to the binary geometry can contaminate the phase-locked template and degrade the sensitivity to the axion signal if it is not modeled appropriately. Possible suspects are binary systems having significant circumstellar matter associated with time-dependent mass transfer, or stars pulsating in a day-scale period. While we do not discuss such variability in detail, we need careful treatment in modeling polarization signature when we apply these binaries for constraining the axion signal.

\bibliography{ref_updated_v6}

@article{Chen:2019fsq,
    author = "Chen, Yifan and Shu, Jing and Xue, Xiao and Yuan, Qiang and Zhao, Yue",
    title = "{Probing Axions with Event Horizon Telescope Polarimetric Measurements}",
    eprint = "1905.02213",
    archivePrefix = "arXiv",
    primaryClass = "hep-ph",
    doi = "10.1103/PhysRevLett.124.061102",
    journal = "Phys. Rev. Lett.",
    volume = "124",
    number = "6",
    pages = "061102",
    year = "2020"
}

@article{Huang:2025rrb,
    author = "Huang, Qiu-Ju and Wang, Bao and Wei, Jun-Jie and Wu, Xue-Feng",
    title = "{Hunting for Extragalactic Axion-like Dark Matter in a Decade-long Blazar Optical Polarimetry}",
    eprint = "2511.05839",
    archivePrefix = "arXiv",
    primaryClass = "astro-ph.HE",
    journal = "",
    month = "11",
    year = "2025"
}

@article{Wang:2024sdz,
    author = "Wang, Bao and Yang, Xuan and Wei, Jun-Jie and Zhang, Song-Bo and Wu, Xue-Feng",
    title = "{Detecting extragalactic axion-like dark matter with polarization measurements of fast radio bursts}",
    eprint = "2402.00473",
    archivePrefix = "arXiv",
    primaryClass = "astro-ph.HE",
    doi = "10.1038/s42005-025-02045-w",
    journal = "Commun. Phys.",
    volume = "8",
    number = "1",
    pages = "130",
    year = "2025"
}

@article{Davydov:2023hgw,
    author = "Davydov, Daniil and Libanov, Alexander",
    title = "{Constraints on axionlike ultralight dark matter from observations of the HL Tauri protoplanetary disk}",
    eprint = "2312.03926",
    archivePrefix = "arXiv",
    primaryClass = "hep-ph",
    doi = "10.1103/PhysRevD.110.103022",
    journal = "Phys. Rev. D",
    volume = "110",
    number = "10",
    pages = "103022",
    year = "2024"
}

@article{POLARBEAR:2025djl,
    author = "Adkins, Tylor and others",
    collaboration = "POLARBEAR",
    title = "{Constraints on the polarization angle oscillations of the Crab Nebula with the Simons Array and its applications to the search for axionlike particles}",
    eprint = "2512.18882",
    archivePrefix = "arXiv",
    primaryClass = "astro-ph.CO",
    doi = "10.1103/9wqc-tgzq",
    journal = "Phys. Rev. D",
    volume = "113",
    number = "4",
    pages = "043044",
    year = "2026"
}

@article{Carroll:1989vb,
    author = "Carroll, Sean M. and Field, George B. and Jackiw, Roman",
    title = "{Limits on a Lorentz and Parity Violating Modification of Electrodynamics}",
    reportNumber = "MIT-CTP-1782",
    doi = "10.1103/PhysRevD.41.1231",
    journal = "Phys. Rev. D",
    volume = "41",
    pages = "1231",
    year = "1990"
}

@article{Harari:1992ea,
    author = "Harari, Diego and Sikivie, Pierre",
    title = "{Effects of a Nambu-Goldstone boson on the polarization of radio galaxies and the cosmic microwave background}",
    reportNumber = "UFIFT-HEP-92-9",
    doi = "10.1016/0370-2693(92)91363-E",
    journal = "Phys. Lett. B",
    volume = "289",
    pages = "67--72",
    year = "1992"
}

@article{Hui:2016ltb,
    author = "Hui, Lam and Ostriker, Jeremiah P. and Tremaine, Scott and Witten, Edward",
    title = "{Ultralight scalars as cosmological dark matter}",
    eprint = "1610.08297",
    archivePrefix = "arXiv",
    primaryClass = "astro-ph.CO",
    doi = "10.1103/PhysRevD.95.043541",
    journal = "Phys. Rev. D",
    volume = "95",
    number = "4",
    pages = "043541",
    year = "2017"
}

@INPROCEEDINGS{2015ASPC..496..164S,
       author = {{Southworth}, J.},
        title = "{DEBCat: A Catalog of Detached Eclipsing Binary Stars}",
    booktitle = {Living Together: Planets, Host Stars and Binaries},
         year = 2015,
       editor = {{Rucinski}, S.~M. and {Torres}, G. and {Zejda}, M.},
       series = {Astronomical Society of the Pacific Conference Series},
       volume = {496},
        month = jul,
        pages = {164},
          doi = {10.48550/arXiv.1411.1219},
archivePrefix = {arXiv},
       eprint = {1411.1219},
 primaryClass = {astro-ph.SR},
       adsurl = {https://ui.adsabs.harvard.edu/abs/2015ASPC..496..164S}
}

@ARTICLE{2019NatAs...3..636B,
       author = {{Bailey}, Jeremy and {Cotton}, Daniel V. and {Kedziora-Chudczer}, Lucyna and {De Horta}, Ain and {Maybour}, Darren},
        title = "{Polarized reflected light from the Spica binary system}",
      journal = {Nature Astronomy},
         year = 2019,
        month = apr,
       volume = {3},
        pages = {636-641},
          doi = {10.1038/s41550-019-0738-7},
       adsurl = {https://ui.adsabs.harvard.edu/abs/2019NatAs...3..636B}
}

@article{Preskill:1982cy,
    author = "Preskill, John and Wise, Mark B. and Wilczek, Frank",
    editor = "Srednicki, M. A.",
    title = "{Cosmology of the Invisible Axion}",
    reportNumber = "HUTP-82-A048, NSF-ITP-82-103",
    doi = "10.1016/0370-2693(83)90637-8",
    journal = "Phys. Lett. B",
    volume = "120",
    pages = "127--132",
    year = "1983"
}

@article{Wilczek:1977pj,
    author = "Wilczek, Frank",
    title = "{Problem of Strong  $P$  and  $T$  Invariance in the Presence of Instantons}",
    reportNumber = "Print-77-0939 (COLUMBIA)",
    doi = "10.1103/PhysRevLett.40.279",
    journal = "Phys. Rev. Lett.",
    volume = "40",
    pages = "279--282",
    year = "1978"
}

@article{Weinberg:1977ma,
    author = "Weinberg, Steven",
    title = "{A New Light Boson?}",
    reportNumber = "HUTP-77/A074",
    doi = "10.1103/PhysRevLett.40.223",
    journal = "Phys. Rev. Lett.",
    volume = "40",
    pages = "223--226",
    year = "1978"
}

@article{Arvanitaki:2009fg,
    author = "Arvanitaki, Asimina and Dimopoulos, Savas and Dubovsky, Sergei and Kaloper, Nemanja and March-Russell, John",
    title = "{String Axiverse}",
    eprint = "0905.4720",
    archivePrefix = "arXiv",
    primaryClass = "hep-th",
    doi = "10.1103/PhysRevD.81.123530",
    journal = "Phys. Rev. D",
    volume = "81",
    pages = "123530",
    year = "2010"
}

@article{Svrcek:2006yi,
    author = "Svrcek, Peter and Witten, Edward",
    title = "{Axions In String Theory}",
    eprint = "hep-th/0605206",
    archivePrefix = "arXiv",
    reportNumber = "SLAC-PUB-11894",
    doi = "10.1088/1126-6708/2006/06/051",
    journal = "JHEP",
    volume = "06",
    pages = "051",
    year = "2006"
}

@article{Peccei:1977hh,
    author = "Peccei, R. D. and Quinn, Helen R.",
    title = "{CP Conservation in the Presence of Instantons}",
    reportNumber = "ITP-568-STANFORD",
    doi = "10.1103/PhysRevLett.38.1440",
    journal = "Phys. Rev. Lett.",
    volume = "38",
    pages = "1440--1443",
    year = "1977"
}

@article{Abbott:1982af,
    author = "Abbott, L. F. and Sikivie, P.",
    editor = "Srednicki, M. A.",
    title = "{A Cosmological Bound on the Invisible Axion}",
    reportNumber = "PRINT-82-0695 (BRANDEIS)",
    doi = "10.1016/0370-2693(83)90638-X",
    journal = "Phys. Lett. B",
    volume = "120",
    pages = "133--136",
    year = "1983"
}

@article{Dine:1982ah,
    author = "Dine, Michael and Fischler, Willy",
    editor = "Srednicki, M. A.",
    title = "{The Not So Harmless Axion}",
    reportNumber = "UPR-0201T",
    doi = "10.1016/0370-2693(83)90639-1",
    journal = "Phys. Lett. B",
    volume = "120",
    pages = "137--141",
    year = "1983"
}

@article{Yuwen:2026zjk,
    author = "Yuwen, Zi-Yan and Sarkis, Michael and Ma, Yin-Zhe and Liu, Tao and Ren, Jing and Weltevrede, Patrick and Xue, Xiao",
    title = "{The MeerKAT Thousand-Pulsar Polarisation Array II: Searches for Ultralight Axion-Like Dark Matter}",
    eprint = "2605.31024",
    journal = "",
    archivePrefix = "arXiv",
    primaryClass = "astro-ph.HE",
    month = "5",
    year = "2026"
}

@article{Narita:2026dvw,
    author = "Narita, Kanako and Fujita, Tomohiro and Tazaki, Ryo and Hatsukade, Bunyo",
    title = "{Searching for Axion-like particle Dark Matter with Time-domain Polarization: Constraints from a protoplanetary disk}",
    eprint = "2602.15611",
    journal = "",
    archivePrefix = "arXiv",
    primaryClass = "astro-ph.CO",
    month = "2",
    year = "2026"
}

@article{Fujita:2018zaj,
    author = "Fujita, Tomohiro and Tazaki, Ryo and Toma, Kenji",
    title = "{Hunting Axion Dark Matter with Protoplanetary Disk Polarimetry}",
    eprint = "1811.03525",
    archivePrefix = "arXiv",
    primaryClass = "astro-ph.CO",
    doi = "10.1103/PhysRevLett.122.191101",
    journal = "Phys. Rev. Lett.",
    volume = "122",
    number = "19",
    pages = "191101",
    year = "2019"
}

@article{cotton2020phase,
  title={Phase-locked polarization by photospheric reflection in the semidetached eclipsing binary $\mu$1 Sco},
  author={Cotton, Daniel V and Bailey, Jeremy and Kedziora-Chudczer, Lucyna and De Horta, Ain},
  journal={Monthly Notices of the Royal Astronomical Society},
  volume={497},
  number={2},
  pages={2175--2189},
  year={2020},
  publisher={Oxford University Press}
}

@BOOK{1960ratr.book.....C,
    author={Chandrasekhar, Subrahmanyan},
  title     = {Radiative Transfer},
  publisher = {Dover Publications},
  address   = {New York},
  year      = {1960}
}

@ARTICLE{1978A&A....68..415B,
       author = {{Brown}, J.~C. and {McLean}, I.~S. and {Emslie}, A.~G.},
        title = "{Polarisation by Thomson scattering in optically thin stellar envelopes. II. Binary and multiple star envelopes and the determination of binary inclinations.}",
      journal = {Astron. Astrophys.},
         year = 1978,
        month = aug,
       volume = {68},
        pages = {415-427},
       adsurl = {https://ui.adsabs.harvard.edu/abs/1978A&A....68..415B}
}

@article{Chigusa:2019,
    author = "Chigusa, So and Moroi, Takeo and Nakayama, Kazunori",
    title = "{Signals of axion like dark matter in time dependent polarization of light}",
    eprint = "1911.09850",
    archivePrefix = "arXiv",
    primaryClass = "hep-ph",
    doi = "10.1016/j.physletb.2020.135288",
    journal = "Phys. Lett. B",
    volume = "803",
    pages = "135288",
    year = "2020"
}

@article{Fedderke:2019,
    author = "Fedderke, Michael A. and Graham, Peter W. and Rajendran, Surjeet",
    title = "{Axion Dark Matter Detection with CMB Polarization}",
    eprint = "1903.02666",
    archivePrefix = "arXiv",
    primaryClass = "astro-ph.CO",
    doi = "10.1103/PhysRevD.100.015040",
    journal = "Phys. Rev. D",
    volume = "100",
    number = "1",
    pages = "015040",
    year = "2019"
}

@article{Ferguson:2022,
    author = "Ferguson, K. R. and others",
    collaboration = "SPT-3G",
    title = "{Searching for axionlike time-dependent cosmic birefringence with data from SPT-3G}",
    eprint = "2203.16567",
    archivePrefix = "arXiv",
    primaryClass = "astro-ph.CO",
    doi = "10.1103/PhysRevD.106.042011",
    journal = "Phys. Rev. D",
    volume = "106",
    number = "4",
    pages = "042011",
    year = "2022"
}

@article{BICEP:2021,
    author = "Ade, P. A. R. and others",
    title = "{BICEP / Keck XIV: Improved constraints on axion-like polarization oscillations in the cosmic microwave background}",
    eprint = "2108.03316",
    archivePrefix = "arXiv",
    primaryClass = "astro-ph.CO",
    journal = "Phys. Rev. D",
    volume = "105",
    number = "2",
    pages = "022006",
    year = "2022",
    doi = "10.1103/PhysRevD.105.022006"
}

@article{Castillo:2022,
    author = "Castillo, Andr{\'e}s and Martin-Camalich, Jorge and Terol-Calvo, Jorge and Blas, Diego and Caputo, Andrea and Santos, Ricardo Tanaus{\'u} G{\'e}nova and Sberna, Laura and Peel, Michael and Rubi{\~n}o-Mart{\'\i}n, Jose Alberto",
    title = "{Searching for dark-matter waves with PPTA and QUIJOTE pulsar polarimetry}",
    eprint = "2201.03422",
    archivePrefix = "arXiv",
    primaryClass = "astro-ph.CO",
    doi = "10.1088/1475-7516/2022/06/014",
    journal = "JCAP",
    volume = "06",
    number = "06",
    pages = "014",
    year = "2022"
}

@article{EPTA:2024gxu,
    author = "Porayko, N. K. and others",
    collaboration = "EPTA",
    title = "{Searches for signatures of ultralight axion dark matter in polarimetry data of the European Pulsar Timing Array}",
    eprint = "2412.02232",
    archivePrefix = "arXiv",
    primaryClass = "astro-ph.CO",
    doi = "10.1103/PhysRevD.111.062005",
    journal = "Phys. Rev. D",
    volume = "111",
    number = "6",
    pages = "062005",
    year = "2025"
}

@article{Liu:2021zlt,
    author = "Liu, Tao and Lou, Xuzixiang and Ren, Jing",
    title = "{Pulsar Polarization Arrays}",
    eprint = "2111.10615",
    archivePrefix = "arXiv",
    primaryClass = "astro-ph.HE",
    doi = "10.1103/PhysRevLett.130.121401",
    journal = "Phys. Rev. Lett.",
    volume = "130",
    number = "12",
    pages = "121401",
    year = "2023"
}

@article{Bailey:2020HIPPI2,
  author       = {Bailey, Jeremy and Cotton, Daniel V. and Kedziora-Chudczer, Lucyna and De Horta, Ain and Maybour, Darren},
  title        = {{HIPPI-2: a versatile high-precision polarimeter}},
  journal      = {Publications of the Astronomical Society of Australia},
  volume       = {37},
  pages        = {e004},
  year         = {2020},
  doi          = {10.1017/pasa.2019.45},
  eprint       = {1911.02123},
  archivePrefix= {arXiv},
  primaryClass = {astro-ph.IM}
}

@ARTICLE{1978ApJ...221..200R,
       author = {{Rudy}, R.~J. and {Kemp}, J.~C.},
        title = "{A polarimetric determination of binary inclinations: results for five systems}",
      journal = {Astrophys. J.},
         year = 1978,
        month = mar,
       volume = {221},
        pages = {200-210},
          doi = {10.1086/156018},
       adsurl = {https://ui.adsabs.harvard.edu/abs/1978ApJ...221..200R}
}

@misc{OHare:2020AxionLimits,
  author       = {O'Hare, Ciaran},
  title        = {{cajohare/AxionLimits}: {AxionLimits}},
  year         = {2020},
  howpublished = {Zenodo},
  doi          = {10.5281/zenodo.3932430},
  url          = {https://doi.org/10.5281/zenodo.3932430},
  note         = {Version v1.0}
}

@ARTICLE{2022ARAA..60...31K,
       author = {{Kurtz}, Donald W.},
        title = "{Asteroseismology Across the Hertzsprung-Russell Diagram}",
      journal = {Annual Review of Astronomy and Astrophysics},
         year = 2022,
        month = aug,
       volume = {60},
        pages = {31-71},
          doi = {10.1146/annurev-astro-052920-094232},
archivePrefix = {arXiv},
       eprint = {2201.11629},
 primaryClass = {astro-ph.SR},
       adsurl = {https://ui.adsabs.harvard.edu/abs/2022ARA&A..60...31K}
}

@ARTICLE{1997AA...327..620S,
       author = {{Soberman}, G.~E. and {Phinney}, E.~S. and {van den Heuvel}, E.~P.~J.},
        title = "{Stability criteria for mass transfer in binary stellar evolution.}",
      journal = {Astronomy and Astrophysics},
         year = 1997,
        month = nov,
       volume = {327},
        pages = {620-635},
          doi = {10.48550/arXiv.astro-ph/9703016},
archivePrefix = {arXiv},
       eprint = {astro-ph/9703016},
 primaryClass = {astro-ph},
       adsurl = {https://ui.adsabs.harvard.edu/abs/1997A&A...327..620S}
}

\end{document}